\documentclass[a4paper,12pt]{article}
\begin{document}
\large
\begin{center}        \LARGE   \bf
Nonconformal Scalar Field in a  \\
Homogeneous Isotropic Space and
the Method of Hamiltonian Diagonalization
\end{center}
   \begin{center}
  \bf Yu.V.Pavlov{\,}\footnote{\large e-mail: pavlov@ipme.ru} \\
   \end{center}
   \begin{center}
{\em  Institute of Mechanical Engineering, Russian Academy of Sciences, \\
      61 Bolshoy, V.O., St.Petersburg, 199178, Russia }
   \end{center}
\hrule
\begin{abstract}   \normalsize
    {\bf Abstract.}
    The diagonalization of the metrical Hamiltonian of a scalar field with
an arbitrary coupling with a curvature in $N$-dimensional homogeneous
isotropic space is performed.
  The energy spectrum of the corresponding quasiparticles is obtained.
    The energies of the quasiparticles corresponding to the diagonal form
of the canonical Hamiltonian are calculated.
   The modified energy-momentum tensor with the following
properties is constructed.
    It coincides with the metrical energy-momentum tensor for
conformal scalar field. Under its diagonalization the energies
of relevant particles of a nonconformal field coincide to the oscillator
frequencies and the density of such particles created in a nonstationary
metric is finite.
    It is shown that the Hamiltonian calculated with the modified
energy-momentum tensor can be constructed as a canonical Hamiltonian
under the special choice of variables.\\  \\
\noindent
{\it PACS}: 04.62.+v.                      \\
{\it Keywords}:   Scalar field,
               Quantum theory in curved space,
               Particle creation.
\end{abstract}
\hrule
\vspace{9mm}

% ***********************************************************************

{\centering \section{Introduction}}

    Quantum field theory in curved space-time is a deeply worked out
part of theoretical physics now (see monographs \cite{GMM,BD}).
  It is of large interest due to its important applications to cosmology
and astrophysics.
   At the same time in this theory there are several problems
that haven't been finally solved till now.
   One of them is definition of the notion of elementary particle in
curved space-time.
  It is caused by the absence of the group of
symmetries like Poincare group in Minkowsky space.
    The definition of vacuum state and particles states for nonconformal
scalar field is being discussed intensely at present
even in case of homogeneous isotropic space \cite{Diag,Redmount99,Lindig}.
    As a consequence of various definitions of vacuum states we have
variety of calculated quantum characteristics of nonconformal scalar field
in curved space.

    In works \cite{Pv,BLM} it was shown that in the case of arbitrary
coupling of scalar field with curvature additional nonconformal
contributions are dominant in vacuum averages of energy-momentum tensor.
   It should be also mentioned that the investigation of nonconformal scalar
field is not of independent interest only. This investigation is
caused by impossibility of preservation of conformal invariation
of effective action and usual action in the case of interacting
quantized field \cite{BD}.

    In the definition of vacuum and in the formulation of the problem
of particles creation in curved space-time two following approaches
are being widely used:
the Hamiltonian diagonalization procedure \cite{GMM} offered in
\cite{G69,G71} and the so-called "adiabatic" procedure \cite{BD}
offered in \cite{Parker}.
    Supposing that a quantum of energy corresponds to a particle then
observation of particles at some moment (according to quantum mechanics)
means to find the Hamiltonian eigenstate.
   It is taken into account automatically in the diagonalization approach.
    Making use of Hamiltonian built according to the
metrical energy-momentum tensor, having been successful for the conformal
case~(see \cite{GMM}).
    But for nonconformal scalar field in diagonalization
approach it led to difficulties, connected
with infinite of quasiparticles creation density~\cite{Fulling}.
    At the same time essential problems and ambiguities take place in
adiabatic approach also~\cite{Lindig}.

     In this paper the procedure of diagonalization for scalar field with
arbitrary coupling with curvature in $N$-dimensional homogeneous
isotropic space is investigated.
     Carrying out calculations in $N$-dimensional space-time while finding
vacuum state and defining particles states is necessary in
particular when the dimensional regularization is used in further.
    In section 2 necessary information is given and nonconformal scalar
field quantizing in $N$-dimensional homogeneous isotropic space is
performed.
    In section 3 the metrical Hamiltonian diagonalization of nonconformal
scalar field is carried out.
   The energies of corresponding quasiparticles are calculated and
conditions connected with demand of the metrical Hamiltonian
diagonality are investigated.
    In section 4 the diagonalization approach is used for the canonical
Hamiltonian, the modified energy-momentum tensor is built so that
the quasiparticles from diagonalization of corresponding Hamiltonian
have energies coinciding with the oscillator frequency of the wave equation.
    It is proved that the density of particles being created in
nonstationary metric is finite.
    In section 5 it is shown that the Hamiltonian obtained from the modified
energy-momentum tensor can be built as canonical  one under a certain
choice of canonical variables and the results of performed
investigation are summarized.

    The system of units in which $\hbar =c=1$ is using in the paper.
\\

{\centering \section{Quantizing of Scalar Field in Homogeneous Isotropic
Space}}

    We will consider the complex scalar field $\phi(x)$ of mass $m$ obeing
the equation
\begin{equation}
 ({\nabla}_i {\nabla}^i + \xi \, R +m^2)  \, \phi(x)=0 \, , \label{1}
\end{equation}
    where ${\nabla}_i$ is covariant derivative, $R$ is the scalar curvature,
$x=(t,{\bf x})$, \
$\xi$ is coupling constant.
    The value $ \xi =\xi_c =(N-2)/\,[\,4\,(N-1)] $
correspond to conformal coupling in space-time of dimension $N$
($ \xi_c=1/6 $ if $N=4$).
    The equation (\ref{1}) is conformaly
invariant if $\xi=\xi_c $ and $m=0$.
   The value $\xi=0$  reduces to the case of minimal coupling.

   The metric of $N$-dimensional homogeneous isotropic space-time is
\begin{equation}
 ds^2=g_{ik}dx^i\,dx^k=dt^2-a^2(t)\,dl^2=a^2(\eta)\,(d{\eta}^2 - dl^2) \,,
 \label{2}
\end{equation}
    where $dl^2=\gamma_{\alpha \beta} \,dx^\alpha \,dx^\beta $
is the metric of $N-1$ dimensional space with the constant curvature
$K=0, \pm 1$.

    The equation (\ref{1}) can be received by varying the action with
Lagrangian density
\begin{equation}
L(x)=\sqrt{|g|}\ [\,g^{ik}\partial_i\phi^*\partial_k\phi -(m^2+\xi R)
\phi^* \phi \,] \, ,
\label{3}
\end{equation}
    where $g=det(g_{ik})$.

    Canonical energy-momentum tensor of scalar field is
\begin{equation}
T_{ik}^{can}=\partial_i\phi^* \partial_k\phi+
\partial_k\phi^* \partial_i\phi-g_{ik} |g|^{-1/2}L(x) \, .
\label{4}
\end{equation}

     The metrical energy-momentum tensor which can be got varying the action
in $g_{ik}$ have a form~\cite{ChT}:
\begin{equation}
T_{ik}=T_{ik}^{can}-2 \xi [R_{ik}+\nabla_i\nabla_k-g_{ik}\nabla_j\nabla^j ]
\, \phi^* \phi \, ,
\label{5}
\end{equation}
    where $R_{ik} $ is Ricci tensor.

    In the metric (\ref{2}) the equation (\ref{1}) is
\begin{equation}
\phi''+(N-2) \left(\frac{a'}{a}\right) \phi'-\Delta_{N-1}\,\phi+
(m^2+\xi R)\,a^2\phi=0 \, ,
\label{6}
\end{equation}
    where $\Delta_{N-1}$  is the Laplace-Beltrami operator in
$N-1$ dimension space, the prime denotes the derivative with
conformal time $\eta$.

    For the function $\tilde{\phi}= a^{(N-2)/\,2}\phi $
the equation (\ref{6}) has the form without the first derivative with time
\begin{equation}
\tilde{\phi}'' -\Delta_{N-1}\,\tilde{\phi} +
\left(m^2 a^2-\Delta\xi a^2 R+ ((N-2)/\,2)^2 \,K \right)\,\tilde{\phi}=0 \,,
\label{7}
\end{equation}
    where  $ \Delta \xi=\xi_c-\xi $. \
The variables in the equations (\ref{6}), (\ref{7}) can be separated.
Namely, for
$ \tilde{\phi}= g_\lambda (\eta) \Phi_J({\bf x}) $   we have
\begin{equation}
g_\lambda''(\eta)+\Omega^2(\eta)\,g_\lambda(\eta)=0 \,,
\label{8}
\end{equation}
\begin{equation}
\Delta_{N-1}\,\Phi_J=-(\lambda^2-((N-2)/\,2)^2\,K )\,\Phi_J \,,
\label{9}
\end{equation}
    where $\Omega(\eta) $ is the oscillator frequency
\begin{equation}
\Omega^2(\eta)=m^2 a^2 + \lambda^2 -\Delta \xi\,a^2 R \,,
\label{10}
\end{equation}
    $J$ is a set of indexes (quantum numbers), numbering the eigenfunctions
of the Laplace-Beltrami operator. It should be noted that the eigenvalue
of the operator $-\Delta_{N-1} $ are not negative and we have the
inequality              \\
$ \lambda^2-((N-2)/\,2)^2\,K \ge 0 $\,.

    For quantization we are decomposing the field $ \tilde{\phi}(x) $
by the complete set of the equation (\ref{7}) solutions
\begin{equation}
\tilde{\phi}(x)=\int d\mu(J)\,\biggl[\,\tilde{\phi}^{(-)}_{\bar{J}}
\,a^{(-)}_{\bar{J}} + \tilde{\phi}^{(+)}_{J}\,a^{(+)}_{J}\,\biggr] \ ,
\label{11}
\end{equation}
    where $ d\mu (J) $ is the measure in the space of the Laplace-Beltrami
$ \Delta_{N-1} $ eigenvalues
\begin{equation}
\tilde{\phi}^{(+)}_J (x)=\frac{1}{\sqrt{2}}\,
g_\lambda(\eta)\,\Phi^*_J({\bf x}) \ ,
\ \  \ \
\tilde{\phi}^{(-)}_{\bar{J}} (x)=\frac{1}{\sqrt{2}}\,
g^*_\lambda(\eta)\,\Phi_{\bar{J}}({\bf x}) \ ,
\label{12}
\end{equation}
    $\Phi_J({\bf x}) $ is orthonormal eigenfunctions of $\Delta_{N-1}$
operator,
$\bar{J} $ is a set of quantum numbers of the function complex conjugated
to the function $\Phi_J $. \
In  $N-1$ dimensional spherical coordinates for
$J=\{\lambda \,, l\,,\ldots \, , m \} \ $    we have
$\bar{J}\!~=~\!\{\lambda \,, l\,,\ldots \, , - m \}  $.

    Substituting the decomposition (\ref{11}) in the equation for
conservative charge
\begin{equation}
Q=i \int \limits_{\Sigma}\left(\tilde{\phi}^* \partial_0\tilde{\phi}
- (\partial_0\tilde{\phi}^* )\,\tilde{\phi}\,\right)
\sqrt{\gamma}\,d^{N-1}x \,,
\label{13}
\end{equation}
    where $\gamma=det(\gamma_{\alpha \beta})$,
    $\Sigma $ is a space-like hypersurface
$\eta\!=\!\mbox{const}$, under the condition of normalization
\begin{equation}
g_\lambda\,g_\lambda^{* \prime } -
g_\lambda^\prime\,g_\lambda^* = -2 i \ ,
\label{norm}
\end{equation}
    we have
\begin{equation}
Q = \int\,d \mu (J) \, \left(\stackrel{*}{a}\!{\!}^{(+)}_J a^{(-)}_J -
\stackrel{*}{a}\!{\!}^{(-)}_{\bar{J}} a^{(+)}_{\bar{J}} \right) \ .
\label{15}
\end{equation}

    The metrical Hamiltonian can be expressed through the metrical
energy-momentum tensor (\ref{5}) by the equation~\cite{GMM}:
\begin{eqnarray}
H(\eta)=\int \limits_\Sigma \zeta^i \, T_{ik}(x)\,d \sigma^k = \!\!
\int \limits_{\eta={\rm const}}\!\!\! \zeta^0 \,T_{00}(x) \,
g^{00}\sqrt{|g|} \ d^{N-1}x & = &       \nonumber   \\
= a^{N-2}(\eta) \!\! \int \limits_{\eta={\rm const}}\!\!\!T_{00}(x) \,
\sqrt{\gamma}\ d^{N-1}x \, ,\phantom{xxxxxxx}   & &
\label{16}
\end{eqnarray}
    where $(\zeta^i)=(1, 0 , \ldots, 0) $ is the time-like conformal
Killing vector.

    The quantization is realized by the commutation relations
\begin{equation}
\left[a_J^{(-)}, \ \stackrel{*}{a}\!{\!}_{J'}^{(+)}\right]=
\left[\stackrel{*}{a}\!{\!}_J^{(-)}, \ a_{J'}^{(+)}\right]=\delta_{JJ'}  \ ,
\ \ \ \left[a_J^{(\pm)}, \ a_{J'}^{(\pm)}\right]=
\left[\stackrel{*}{a}\!{\!}_J^{(\pm)}, \ \stackrel{*}{a}\!{\!}_{J'}^{(\pm)}
\right]=0
\,.  \label{17}
\end{equation}
    The Hamiltonian (\ref{16}) can be written through
$ a_J^{(\pm)} , \ \stackrel{*}{a}\!{\!}_J^{(\pm)} $ operators in such form
\begin{eqnarray}
H(\eta)&=&\int\, d\mu(J)  \biggl\{ E_J(\eta) \,
\left(\stackrel{*}{a}\!{\!}^{(+)}_J a^{(-)}_J +
\stackrel{*}{a}\!{\!}^{(-)}_{\bar{J}} a^{(+)}_{\bar{J}} \right) +
  \nonumber \\
&+& F_J(\eta) \, \stackrel{*}{a}\!{\!}^{(+)}_J a^{(+)}_{\bar{J}} +
F^*_J(\eta)\, \stackrel{*}{a}\!{\!}^{(-)}_{\bar{J}} a^{(-)}_J  \biggr\} \,,
\label{H}
\end{eqnarray}
    where
\begin{equation}
E_J(\eta)=\frac{1}{2}\left\{\,|g_\lambda'|^2+
D_\lambda(\eta)\,|g_\lambda|^2 -
Q(\eta)\,(|g_\lambda|^2)' \, \right\}  \,,
\label{EJ}
\end{equation}
\begin{equation}
F_J(\eta)=\frac{(-1)^m}{2}\left\{{g_\lambda'}^2+
D_\lambda(\eta)\,g_\lambda^2 -
Q(\eta)\,(g_\lambda^2)'   \, \right\}  \,,
\label{FJ}
\end{equation}
\begin{equation}
D_\lambda(\eta)=m^2 a^2+\lambda^2+\Delta\xi\,(N-1)\,(N-2)\,(c^2-K) \,,
\label{21}
\end{equation}
\begin{equation}
Q(\eta)=\Delta \xi \, 2 \,(N-1) \, c    \,,
\label{22}
\end{equation}
   $c=a'(\eta)/ a(\eta)$ .
\\

\section{The Diagonalization of the Metrical Hamiltonian}

\hspace{\parindent}
    The metrical Hamiltonian (\ref{H}) will be diagonal at time moment
$\eta_0 $ in operators
$ \stackrel{*}{a}\!{\!}^{(\pm)}_J, \  a^{(\pm)}_J $,
which in this case are the creation and annihilation operators of particles
and antiparticles under  $F_J(\eta_0)=0$.
    Using (\ref{FJ}) -- (\ref{22}) we can check that this condition
is consistent with normalization (\ref{norm}) only under
$p^2_\lambda(\eta_0)>0 $,
  where
\begin{equation}
p_\lambda(\eta)=\sqrt{m^2a^2(\eta) + \lambda^2 + 4\Delta\xi \,
(N-1)^2\,(\xi\, c^2 -\xi_c \, K \,)} \,.
\label{23}
\end{equation}

    Requirement of the diagonal form of Hamiltonian at time moment
$\eta_0$, i.e. $F_J(\eta_0)=0$, with the condition of normalization
result in the following initial conditions for the function
$g_\lambda(\eta_0)$
\begin{equation}
g_\lambda'(\eta_0)=( 2 \Delta\xi (N-1)\, c +
i p_\lambda(\eta_0))\, g_\lambda(\eta_0) \,, \ \
|g_\lambda(\eta_0)|= 1/\sqrt{p_\lambda(\eta_0) } \,.
\label{24}
\end{equation}
    The state of vacuum $|\,0\!>$ corresponding to (\ref{24})
is defined in a standard way
\begin{equation}
a^{(-)}_J|\,0\!> \,=\, \stackrel{*}{a}\!{\!}^{(-)}_J|\,0\!  >\, = 0 \,.
\label{25}
\end{equation}

    For arbitrary time moment $\eta $ we will diagonalize the Hamiltonian
in terms of
$ \stackrel{*}{b}\!{\!}^{(\pm)}_J, \  b^{(\pm)}_J $ operators which
    connected with
$ \stackrel{*}{a}\!{\!}^{(\pm)}_J, \  a^{(\pm)}_J $
    by the time-dependent Bogolyubov transformations:
\begin{equation}
\left\{  \begin{array}{c}
a_J^{(-)}=\alpha^*_J(\eta) \,b^{(-)}_J(\eta)-
(-1)^m \beta_J(\eta)\, b^{(+)}_{\bar{J}}(\eta) \,,  \\[3mm]
\stackrel{*}{a}\!{\!}_J^{(-)}=\alpha^*_J(\eta) \,
\stackrel{*}{b}\!{\!}^{(-)}_J\!(\eta)-
(-1)^m \beta_J(\eta) \,\stackrel{*}{b}\!{\!}^{(+)}_{\bar{J}}\!(\eta) \,,
\end{array} \right.
\label{26}
\end{equation}
  where $\alpha_J(\eta), \ \beta_J(\eta) $ is the functions satisfy
the initial conditions         \\ $|\alpha_J(\eta_0)|=1, \
\beta_J(\eta_0)=0 $ and the identity
\begin{equation}
|\alpha_J(\eta)|^2-|\beta_J(\eta)|^2=1       \,.
\label{27}
\end{equation}
(In homogeneous and isotropic space
$\alpha_J =\alpha_\lambda , \ \beta_J=\beta_\lambda $ \cite{GMM}).

    Let's substitute decomposition (\ref{26}) in (\ref{H}).
If we demand coefficients before nondiagonal terms
$\stackrel{*}{b}\!{\!}^{(\pm)}_J b^{(\pm)}_J $ equal to 0 we'll
come to equation
\begin{equation}
2 (-1)^{m+1} \alpha_J \beta_J E_J+ F_J \alpha_J^2+F_J^* \beta_J^2=0 \,.
\label{28}
\end{equation}
    It can be shown that the condition (\ref{28}) is consistent with
normalization (\ref{norm}) only if $p^2_\lambda(\eta)>0$.
    In that case
\begin{equation}
|\beta_J|^2=E_J/\,(2p_\lambda)-1/2=|F_J|^2/\,(2 p_\lambda \,
(E_J+p_\lambda ))    \,.
\label{29}
\end{equation}
    Concluding (\ref{29}) we took into account the equation
that could be easily checked:
\begin{equation}
E_J^2- |F_J|^2 = p^2_\lambda(\eta) \!\cdot\!
\left[  - ( g_\lambda\,g_\lambda^{* \prime } -
g_\lambda^\prime\,g_\lambda^*\, )^2 /\,4\,\right] \,.
\label{30}
\end{equation}
    (The multiplier in square brackets equals to 1 under the normalization
condition (\ref{norm})\,).

    In the case of (\ref{28}) and $ p^2_\lambda(\eta)>0 $
the metrical Hamiltonian takes the form
\begin{equation}
H(\eta) =\int d\mu(J) \,p_\lambda(\eta) \,
\left(\,\stackrel{*}{b}\!{\!}^{(+)}_J b^{(-)}_J +
\stackrel{*}{b}\!{\!}^{(-)}_{\bar{J}} b^{(+)}_{\bar{J}}\, \right) \,.
\label{31}
\end{equation}
    So $p_\lambda(\eta) $ has the meaning of energy of quasiparticles
corresponding to the diagonal form of the metrical Hamiltonian.
    For the 4 dimensional space-time the equation (\ref{23})
corresponds to energy values obtained in \cite{Diag} and~\cite{CF86}\,.

    Quasiparticles' energy $p_\lambda(\eta) $ differ from oscillator
frequency $\Omega(\eta) $ of the wave equation for nonconformal field.
   It leads to a series of difficulties.
Thus the conditions $p^2_\lambda(\eta)>0 $ and $\ \Omega^2(\eta)>0 $
may be in contradiction for nonconformal field in some case.
For example, in the case of quasi-euclidean space ($K=0 $) and
zero-mass field the condition $p^2_\lambda(\eta)>0 $
(with arbitrary~$\lambda$) reduces to
$\ \xi \in [\, 0 , \,\xi_c \,] $.
But if $\xi<\xi_c \,, \ m=0 $ and  $R>0 $ for low $\lambda $
we have $\ \Omega^2(\eta)< 0 $.

    It should be noted that for $p^2_\lambda(\eta)<0 $ the condition
of diagonalization reduces to vanishing of norm, energy and charge of
the state with $\phi(x)\ne 0 $,
that hasn't any physical foundation.

    The vacuum state defined by equations
\begin{equation}
b^{(-)}_J|\,0_\eta\!> \,=\, \stackrel{*}{b}\!{\!}^{(-)}_J|\,0_\eta\!>\,
= 0 \,,
\label{32}
\end{equation}
   depends on time in nonstationary metric.
Under initial conditions (\ref{24})
$ b^{(\pm)}_J(\eta_0)=a^{(\pm)}_J $ and $|\,0_{\eta_0}\!>=|\,0\!> $.
In the Heisenberg representation the state $|\,0\!>$ is vacuum at time
moment $\eta_0$ but it isn't vacuum if $\eta\ne\eta_0$.
It has $|\beta_J(\eta)|^2$ pairs of quasiparticles corresponding to
operators
$\stackrel{*}{b}\!{\!}^{(\pm)}_J\,,   b^{(\pm)}_J$ \cite{GMM}
in each mode.
    The number of the created pairs of quasiparticles in the unit of space
volume (for $N=4$) is
\begin{equation}
n(\eta)=\frac{1}{2 \pi^2 a^3(\eta)}
\int d\mu(J)\,|\beta_\lambda(\eta)|^2 \,.
\label{33}
\end{equation}
   For the metrical Hamiltonian of nonconformal field the density of
quasiparticles created in nonstationary metric (\ref{33}) is
infinite~\cite{Fulling}.
    It follows from asymptotic behaviour in $\lambda \to \infty $
of equation's (\ref{8}) solutions and equation~(\ref{29}).

\section{Diagonalization of the Canonical Ha\-mil\-tonian and the
Modified Energy-Momentum Tensor}

\hspace{\parindent}
    The Hamiltonian corresponding to canonical energy-momentum
tensor~(\ref{4}) can be written in form (\ref{H}) with change of
(\ref{21}) and (\ref{22}) by
\begin{eqnarray}
D_\lambda(\eta)&=&m^2 a^2+\lambda^2+(N-1)\,(N-2)\,\left(\,(\xi+\xi_c)\,
c^2 - \Delta \xi\,K \, \right) + \nonumber  \\*
&& \phantom{xxxxxxxx} +2\,\xi\,(N-1)\,c'  \,,
\label{34}
\end{eqnarray}
\begin{equation}
Q(\eta)=c\, (N-2)/\,2    \,.
\label{35}
\end{equation}
    That's why the procedure of the canonical Hamiltonian diagonalization
can be carried out as in the case of the metrical Hamiltonian
(see section 3) with new functions
$D_\lambda(\eta)$ and $Q(\eta)$.
One can be convinced that the requirement of the Hamiltonian diagonality
is consistent now to normalization (\ref{norm})
only in the case of $p^2_{can,\lambda}(\eta)>0 $, where
\begin{equation}
p_{can,\lambda}(\eta)=\sqrt{(m^2+\xi\,R)\,a^2(\eta) + \lambda^2 -
((N-2)/\,2)^2\,K} \,.
\label{36}
\end{equation}
    The initial conditions corresponding to the diagonal form of the
Hamiltonian at time $\eta=\eta_0 $ if $p^2_{can,\lambda}(\eta_0)>0 $
and with normalization (\ref{norm}) can be written in the form
\begin{equation}
g_\lambda'(\eta_0)=((N-2)\,c/\,2 +
i\, p_{can,\lambda}(\eta_0))\, g_\lambda(\eta_0) \,, \ \
|g_\lambda(\eta_0)|= 1/\sqrt{p_{can,\lambda}(\eta_0) } \,.
\label{37}
\end{equation}
    Since for the canonical Hamiltonian with
$p^2_{can,\lambda}(\eta)>0 $
the equation~(\ref{31}) with substitution
$p_\lambda \rightarrow p_{can,\lambda}$ \ is correct then functions
$p_{can,\lambda}(\eta) $ mean the energies of quasiparticles  corresponding
to the canonical Hamiltonian diagonal form.
   One should note that $p_{can,\lambda}(\eta) $ energies differ from
oscillator frequency $\Omega(\eta) $  similar to the
case of the metrical Hamiltonian.

    Let us consider the density problem of created in nonstationary metric
pairs of quasiparticles in diagonalization procedure for the canonical
Hamiltonian.
   For asymptotic solutions of equation (\ref{8}) (see \cite{Fed}),
normalized according to (\ref{norm}),
we can obtain from (\ref{EJ}), (\ref{FJ}) and (\ref{34}), (\ref{35})
that $E_J \sim \lambda $ and for nonstationary metric
$|F_J(\eta)| \sim |Q(\eta)| $ in  $\lambda \to \infty $.
Therefore according to (\ref{29}) that is corrected with substitution of
$p_\lambda \rightarrow p_{can,\lambda}$, we have
$|\beta_\lambda|^2 \sim \lambda^{-2} $.
    Consequently, the density of created quasiparticles, proportional to
integral in (\ref{33}), is infinite.

    So, in diagonalization procedure both, for the metrical and the
canonical Hamiltonian in nonconformal scalar field there is a problem
of infinite density of quasiparticles created in nonstationary metric.
     In both cases the energies of corresponding quasiparticles differ
from the oscillator frequency of the wave equation.
    It will be shown below that these difficulties are absent in the case
of the Hamiltonian corresponding to the modified energy-momentum tensor
\begin{equation}
T^{\,mod}_{ik}=T_{ik}^{can}-2 \xi_c\, [R_{ik}+\nabla_i\nabla_k-g_{ik}
\nabla_j\nabla^j ] \, \phi^* \phi \, .
\label{Tmod}
\end{equation}

    From the definition (\ref{Tmod}) it is clear that for conformal scalar
field (i.e. if $\xi=\xi_c $) \  $T^{\,mod}_{ik} $ coincide with metrical
energy-momentum tensor~(\ref{5}).
    The structure of the Hamiltonian constructed by $T^{\,mod}_{ik} $
like (\ref{16}) is
\begin{eqnarray}
H^{mod}(\eta) &=& \int d^{N-1}x\,\sqrt{\gamma} \, \biggl\{
\tilde{\phi}^{* \prime} \tilde{\phi}'
+\gamma^{\alpha \beta}\partial_\alpha\tilde{\phi}^*
\partial_\beta\tilde{\phi}+  \nonumber     \\
&+& \! \Bigl[\, m^2a^2-\Delta \xi\, a^2 R +
\Bigl((N-2)/\,2\Bigr)^2 K\,\Bigr]\,
 \tilde{\phi}^* \tilde{\phi} \, \biggr\} \,.
\label{39}
\end{eqnarray}
    It can be written in form (\ref{H}), but in that case
$ Q(\eta)=0 $ and  $\ D_\lambda(\eta)=\Omega^2(\eta) $.
   Under its diagonalization by
$\stackrel{*}{b}\!{\!}^{(\pm)}_J\,,   b^{(\pm)}_J$,
operators we have (\ref{31}) with change $p_\lambda \rightarrow \Omega $.
Oscillator frequency $\Omega(\eta) $ then coincide with the energy
of corresponding particles.
    The initial conditions  for $g_\lambda(\eta) $ corresponding
to the diagonal form at time moment $\eta_0 $ by operators
$\stackrel{*}{a}\!{\!}^{(\pm)}_J\,,   a^{(\pm)}_J$ (\ref{17}) are
\begin{equation}
g_\lambda'(\eta_0)=i\, \Omega(\eta_0)\, g_\lambda(\eta_0) \,, \ \ \
|g_\lambda(\eta_0)|= 1/\sqrt{\Omega(\eta_0) } \,.
\label{40}
\end{equation}
   They coincide with initial conditions used in \cite{BLM} if
$\arg g_\lambda(\eta_0)=0 $ is fixed.
  In radiation dominated background ($R=0$) they coincides with
conditions used in \cite{Pv, MMSH}.

    Let us show finite density of the created  in nonstationary
metric particles corresponding to the diagonal form of Hamiltonian, built
by $T^{\,mod}_{ik} $.
    It is necessary to find asymptotic in $\lambda \to \infty $
of the function $|\beta_\lambda(\eta)|^2 $.
    The functions $\beta_\lambda(\eta) $ and $\ \alpha_\lambda(\eta) $,
being the solutions (\ref{28}) and satisfying identity (\ref{27}),
can be found in the form
\begin{equation}
\beta_\lambda(\eta)=\frac{i}{2} \frac{e^{i\,\Theta(\eta_0, \eta)}}
{\sqrt{\Omega}}\, \biggl( g'(\eta)-i\,\Omega\, g(\eta)\biggr)  \,,
\label{41}
\end{equation}
\begin{equation}
\alpha_\lambda(\eta)=\frac{i}{2} \frac{e^{i\,\Theta(\eta_0, \eta)}}
{\sqrt{\Omega}}\, \biggl( g^{* \prime}(\eta)-i\,\Omega\, g^*(\eta)
\biggr)  \,,
\label{42}
\end{equation}
    where
$ \Theta(\eta_1, \eta_2) = \int \limits_{\eta_1}^{\eta_2}
\Omega(\eta)\,d\eta $.
  \  In consequence of (\ref{41}--\ref{42}) and equation (\ref{8})
the functions
$s_\lambda(\eta)=|\beta_\lambda(\eta)|^2 $ and
$\ f_\lambda(\eta)=2\,\alpha_\lambda(\eta)\, \beta_\lambda(\eta)
\exp[-2i\,\Theta(\eta_0,\eta)] $
    satisfy to the system of equations:
\begin{equation}
\left\{  \begin{array}{l}
s_\lambda'(\eta)={\displaystyle \frac{\Omega'}{2\,\Omega}}\,
{\rm Re} f_\lambda(\eta)    \,,  \\[3mm]
f_\lambda'(\eta)+2\,i\,\Omega\,f_\lambda(\eta)=
{\displaystyle \frac{\Omega'}{\Omega}} \,(1+2 s_\lambda(\eta)) \,.
\end{array} \right.
\label{43}
\end{equation}
    Taking into account the initial condition
$ s_\lambda(\eta_0)=f_\lambda(\eta_0)=0 $
(as $\beta_\lambda(\eta_0)=0 $) we write the system of differential
equations (\ref{43}) in the equivalent form of the system of Volterra
integral equations
\begin{equation}
f_\lambda(\eta)=\int \limits_{\eta_0}^\eta w(\eta_1)\,
(1+2 s_\lambda(\eta_1))\, \exp[-2\,i\,\Theta(\eta_1,\eta)]\,d\eta_1  \,,
\label{44}
\end{equation}
\begin{equation}
s_\lambda(\eta)=\frac{1}{2}\,\int \limits_{\eta_0}^\eta d\eta_1 \,
w(\eta_1)\, \int \limits_{\eta_0}^{\eta_1} d\eta_2 \,w(\eta_2)\,
(1+2 s_\lambda(\eta_2)) \cos[2\,\Theta(\eta_2,\eta_1)]  \,,
\label{45}
\end{equation}
    where $w(\eta)=\Omega'(\eta)/\,\Omega(\eta) $.
    To find asymptotic $ s_\lambda(\eta) $ let us consider the first
iteration in equation (\ref{45}) and as
$ \Theta(\eta_2, \eta_1 )\to \lambda (\eta_1-\eta_2) $
in $\lambda \to \infty $ we will have
\begin{equation}
s_\lambda(\eta) \approx \frac{1}{4}\,\left|\,\int \limits_{\eta_0}^\eta
w(\eta_1)\,\exp( 2\,i\,\lambda\,\eta_1 )\,d\eta_1 \right|^2 \,.
\label{46}
\end{equation}
    Therefore $s_\lambda \sim \lambda^{-6} $ and integral in (\ref{33})
is finite.

    Thus in this case the density of created particles is finite for
4 dimensional space-time.
    If $t\ll 1/m $ in special case the density of created nonconformal
particle was calculated in \cite{BLM}.

\section{The Modified Hamiltonian as Canonical}

\hspace{\parindent}
    We'll show that the modified Hamiltonian (\ref{39}) can be
obtained in homogeneous isotropic space as canonical under the certain
choice of variables describing scalar field.
    Let's add $N$-divergence $\frac{\partial J^i}{\partial x^i}$,
to the lagrangian density (\ref{3}),
where in $(\eta, {\bf x})$ system of coordinates $\ N$-vector
$\ J^i=(\sqrt{\gamma}\,c\,\tilde{\phi}^*\, \tilde{\phi}\,(N-~\!\! 2)/2,\, 0,
\, \ldots \,, 0) $.
   The movement equations (\ref{1}) are invariant under this addition.
Let us choose
$\tilde{\phi}(x)= a^{(N-2)/\,2}(\eta)\phi(x) $
and $\tilde{\phi}^*(x)$ as the field's variables. In these variables
the equation (\ref{1}) have a form~(\ref{7}).
    From the Lagrangian density
$ L^{\Delta}(x)=L(x)+\frac{\partial J^i}{\partial x^i}$ \ it is follows
that the Hamiltonian density
$h(x)=\tilde{\phi}'\,(\partial L^{\Delta})/(\partial \tilde{\phi}')+
\tilde{\phi}^{* \prime}\,(\partial L^{\Delta})/
(\partial \tilde{\phi}^{* \prime})-L^{\Delta}(x) \ $
    takes the value
\begin{eqnarray}
h(x) &=& \! \sqrt{\gamma} \, \biggl\{ \,
\tilde{\phi}^{* \prime} \tilde{\phi}'
+\gamma^{\alpha \beta}\partial_\alpha\tilde{\phi}^*
\partial_\beta\tilde{\phi}+  \nonumber     \\*
&+& \! \Bigl[\, m^2a^2-\Delta \xi\, a^2 R +
\Bigl((N-2)/\,2\Bigr)^2 K\,\Bigr]\,
 \tilde{\phi}^* \tilde{\phi} \, \biggr\} \,.
\label{47}
\end{eqnarray}
    That's why the Hamiltonian (\ref{39}) is a canonical one for scalar
field if $\tilde{\phi}(x) $ and $\ \tilde{\phi}^*(x) $ are chosen as
the field's variables.

    In the presented work the metrical, canonical and introduced modified
Hamiltonian were investigated.
   It was shown that the density of particles created in nonstationary
homogeneous isotropic space metric is finite only in the case of
modified Hamiltonian (\ref{39}) and the energies of such particles are
equal to the oscillator frequency.

    Modified energy-momentum tensor (\ref{Tmod}) introduced above coincide
with the metrical one for conformal scalar field.
    In homogeneous isotropic space $T_{ik}^{\,mod} $ results in modified
Hamiltonian (\ref{39}) that can be obtained as well as canonical under
the special choice of field's variables.

    One can see that considering a line combination of metrical (\ref{5})
and canonical (\ref{4}) tensors we can certainly get modified tensor
(\ref{Tmod}) if the quasiparticles' energy coincides with oscillator
frequency.
     It should be stressed that metrical energy-momentum tensor can't be
changed to $T_{ik}^{\,mod} $ in the right-hand sides of
Einsteine's equations because $T_{ik}^{\,mod} $ is not covariant
conservation.
     But under corpuscular interpretation of nonconformal scalar field and
when diagonalization procedure is used the Hamiltonian (\ref{39})
constructed by $T_{ik}^{\,mod} $ is preferable in comparison with
the metrical Hamiltonian.

 {\bf Acknowledgments.} The author is grateful to Prof. A.A.Grib and
participants in seminar of A.A.Friedmann Laboratory of  Theoretical
Physics for helpful discussions.

\end{document}